**Effect of Plasma Treatment on the Spontaneous Formation and Morphology of Surface Nanobubbles on Silicon**

Anayet Ullah Siddique and Roseanne Warren*

Department of Mechanical Engineering

University of Utah

1495 E 100 S, 1550 MEK

Salt Lake City, UT 84112, USA

*roseanne.warren@utah.edu

**Abstract**

The formation of gaseous, spherical cap-shaped domains (so-called "surface nanobubbles") at the solid-liquid interface is a topic of fundamental interest due to the possible effects of nanobubbles on surface cleaning, wetting, and nanoscale patterning. This work investigates the spontaneous formation of surface nanobubbles on oxygen plasma-treated Si by PeakForce quantitative nanomechanics (PFQNM) imaging, X-ray photoelectron spectroscopy, and water contact angle measurements. Large quantities of surface nanobubbles with sub-10 nm height and sub-100 nm base width are observed on oxygen plasma-treated Si surfaces that have been "aged" in atmospheric conditions (stored in a plastic wafer container). Several days of aging time are required for surface nanobubbles to form on oxygen plasma-treated Si, at which point the bubbles are remarkably consistent in their properties across samples aged 5-12 days. The presence of surface nanobubbles on plasma-treated aged Si surfaces runs contrary to prior reports of infrequent nanobubble formation on $Si/SiO_x$. Surface characterization supports a theory of post-plasma atmospheric hydrocarbon adsorption inducing changes in Si-wetting behavior to a sufficient extent that nanobubble formation can occur. The results are of importance in Si-based fabrication processes employing oxygen plasma treatment with subsequent DI water immersion.

**Introduction**

Surface nanobubbles are gaseous nanodomains that exist on various substrates immersed in fluid[1]. While surface nanobubbles have many promising applications in, *e.g.,* surface cleaning [2,3], mineral flotation[4], and nanoscale patterning [5–7], they pose problems in certain applications by impeding catalysis or wetting[8] or introducing compressible domains in micro/nanofluidic devices[8–11]. An understanding of the conditions under which surface nanobubbles form is critical to controlling their presence in such applications.

Surface hydrophobicity plays a crucial role in nanobubble formation and stability. Prior research indicates that nanobubbles preferentially form on hydrophobic surfaces, including polystyrene[12], silane-modified Si surfaces[13,14], and gold[15]. On hydrophilic surfaces such as mica or silica, nanobubbles are typically only observed following ethanol-water exchange[16] or under high gas saturation conditions, and their presence is considerably less in terms of both density and volume compared to hydrophobic surfaces[17–19].

Consistent with these observations, Si surfaces immersed in water are generally free of nanobubbles due to the presence of a hydrophilic native oxide[20]. During a typical microfabrication process, Si surfaces undergo multiple wet and/or dry-processing steps that change oxide surface chemistry[21,22]. Among dry cleaning methods, oxygen plasma holds particular significance in effectively removing organic contamination[23]. During plasma treatment, active radicals such as $O_2^+$, $O_2^-$, $O_3$, $O$, $O^+$, $O^-$, and free electrons bombard the surface[24], increasing surface reactivity, dangling bonds, and hydrophilicity. Following plasma treatment, the surface contains a significant number of Si-OH groups. Over time, Si-OH groups are converted to Si-CH$_x$ groups due to the adsorption of hydrocarbons from the atmosphere[25,26], and the Si surface gradually becomes more hydrophobic[27,28]. The significant effect of oxygen plasma on Si hydrophobicity is evident from the numerous applications employing plasma as a pre-treatment to control wetting, including wafer bonding[29,30], large-scale transfer of graphene[24], and micro-electromechanical systems[31].

Despite the importance of plasma treatment and its direct use in manipulating Si hydrophobicity, there have to-date been no direct studies of nanobubble formation on plasma-treated Si. This work provides the first examination of the effect of oxygen plasma treatment on Si nanobubble formation and its evolution in time. The spontaneous formation method was used to produce surface nanobubbles on Si by depositing a drop of deionized water onto the wafer surface. This method was chosen to most closely approximate Si-water exposure conditions that arise during microfabrication. The tendency towards nanobubble formation was evaluated immediately after oxygen plasma treatment and after a period of one to twelve days. Atomic force microscopy (AFM) imaging in PeakForce quantitative nanomechanics (PFQNM) mode was used to quantify bubble density and height profiles over time. X-ray photoelectron spectroscopy (XPS) and water contact angle measurements relate bubble formation to plasma-treated Si's evolving surface hydrophobicity. The results provide important insight into the role of adventitious carbon in promoting nanobubble formation and the potential for nanoscale bubbles to form on plasma-treated Si surfaces when immersed in fluid.

**Materials and Methods**

Sample preparation

As-purchased Si wafers (2-inch, p-type, prime grade) were treated with oxygen plasma using a Technics PE II-A tool operating at 195 W and 300 mTorr ($O_2$ flow rate 31.2 sccm). Wafers were exposed to oxygen plasma for 1 minute. Substrates designated as "plasma-treated fresh Si" surfaces were characterized immediately after treatment *via* water contact angle measurement, XPS, and AFM imaging in air and fluid. Substrates designated as "plasma-treated aged Si" surfaces were prepared by storing plasma-treated fresh Si surfaces for up to twelve days in a sealed plastic container in an ambient environment (21-23º C, 40±10% relative humidity).

Surface characterization

Water contact angles of plasma-treated fresh Si and plasma-treated aged Si surfaces were measured using a contact angle goniometer (model no. 100-00-115, ramé-hart instrument co.) and sessile drop method. Each measurement was repeated at least five times at different surface locations for each substrate and the average result is reported. An X-ray photoelectron spectroscope (Kratos Axis Ultra DLD, monochromatic Al Kα X-ray source) was used to analyze the wafer surface composition. XPS surveys were collected at a fixed analyzer pass energy of 160 eV; narrow scans were acquired at 40 eV pass energy. Charge neutralization was performed for all samples. Binding energies were referenced to the C 1s binding energy of adventitious carbon, which was taken to be 284.5 eV. CasaXPS software was used to analyze all spectral data.

AFM experiments

AFM measurements were conducted using a Dimension Icon AFM with Nanoscope Analysis software operating in PFQNM mode (Bruker, USA). A silicon nitride cantilever (Bruker) with a nominal spring constant of 0.4 N/m and a tip radius of 2 nm was used for imaging in air. A silicon nitride cantilever (DNP-C, Bruker) with a nominal spring constant of 0.24 N/m and a tip radius of 20 nm was used for imaging in fluid. DNP-C is a standard silicon nitride probe used for imaging gaseous domains in fluid [32,33]. In PFQNM mode, the software automatically optimizes the image by adjusting scanning parameters in real time. PFQNM mode determines height using force control such that tip correction is not needed. The deflection sensitivity and spring constant of fluid AFM cantilevers were calibrated *via* the ramp and thermal tuning method on Si surfaces in water. A peak force setpoint of 200-500 pN was applied for imaging.

Fluid AFM experiments were conducted using deionized (DI) water (18.2 MΩ-cm) obtained from a Milli-Q system (Millipore Corp., Boston, MA). To prevent contamination, a glass beaker and glass syringe were used and cleaned with ethanol and pure DI water before use. Prior to each experiment, the AFM liquid cell was rinsed with ethanol, flushed with DI water, and then dried with nitrogen. The fluid cell was then mounted with the AFM head (scanner) and sealed with a silicone O-ring. Substrates were positioned inside the AFM fluid cell, and room-temperature DI water was injected into the system. Sample surfaces were immersed in water for 20-30 min before imaging. Typical scan sizes were 2 µm, and scan time for each image was 8-9 min. All imaging was performed in ambient conditions. Substrates were not treated by any heating process, and only Milli-Q water was used without solvent exchange. All bubbles in fluid AFM images were detected and analyzed using Nanoscope Analysis software. Due to significant errors measuring small bubbles, nanobubbles with a base width greater than 30 nm and a height greater than 2 nm were considered. In total, more than 230 nanobubbles were included in the results measurements.

**Results and Discussion**

AFM Imaging of Plasma-Treated Si in Air and Water

Figure 1 compares the surface morphology of oxygen plasma-treated fresh and aged Si surfaces characterized by PFQNM imaging. The processing step(s) after which each image was obtained are shown schematically in Figure 1a. A representative topographical image is provided for each surface in air and fluid (Figure 1b-e).

Plasma-treated fresh Si is clean and smooth with an RMS roughness in air of 0.1 nm over a 2 μm × 2 μm area (Figure 1b, "Air"). A PFQNM image of plasma-treated fresh Si in fluid displays a smooth surface that closely resembles the image obtained in the air (Figure 1b, "Fluid"). There is no evidence of bubble formation on plasma-treated fresh Si, confirming prior observations that a pristine Si surface does not exhibit any signs of features which are "soft," i.e., deformable in nature[16].

A PFQNM image of plasma-treated aged Si in air shows similar surface morphology as plasma-treated fresh Si in air, with no measurable change in surface roughness (RMS roughness 0.1 nm) (Figure 1c "Air"). In fluid, plasma-treated aged Si shows hemispherical features covering the entire surface in a random pattern (Figure 1c, "Fluid"). The estimated size of these features is ~30-100 nm in base width and ~2-15 nm in height. The size range of these nanodomains and their distribution closely resembles previous observations of surface nanobubbles[17,18,34,35]. It is worth noting that the bubbles seen in Figure 1c are considerably smaller and less densely populated compared to those forming on smooth hydrophobic surfaces[1]. Plasma-treated aged Si in Figure 1c was aged for five days.

Plasma-treated aged Si was once more exposed to oxygen plasma to explore the reversibility of the observed plasma treatment/aging effects. PFQNM measurements in air indicate a slight increase in RMS roughness from 0.1 nm measured previously for plasma-treated fresh Si to 0.2 nm for the re-exposed substrate (Figure 1d, "Air"). Surface nanobubbles do not form on plasma-treated aged Si immediately after a second plasma treatment (Figure 1d, "Fluid"), consistent with the observations of Figure 1b.

Re-exposed plasma-treated Si was then aged for four days. PFQNM measurements in air indicate a consistent RMS roughness of 0.2 nm (Figure 1e, "Air"). Nanobubbles in the size range of tens of nanometers are spontaneously produced on the surface in fluid (Figure 1e, "Fluid"), indicating complete reversibility of the observed plasma treatment/aging effects.

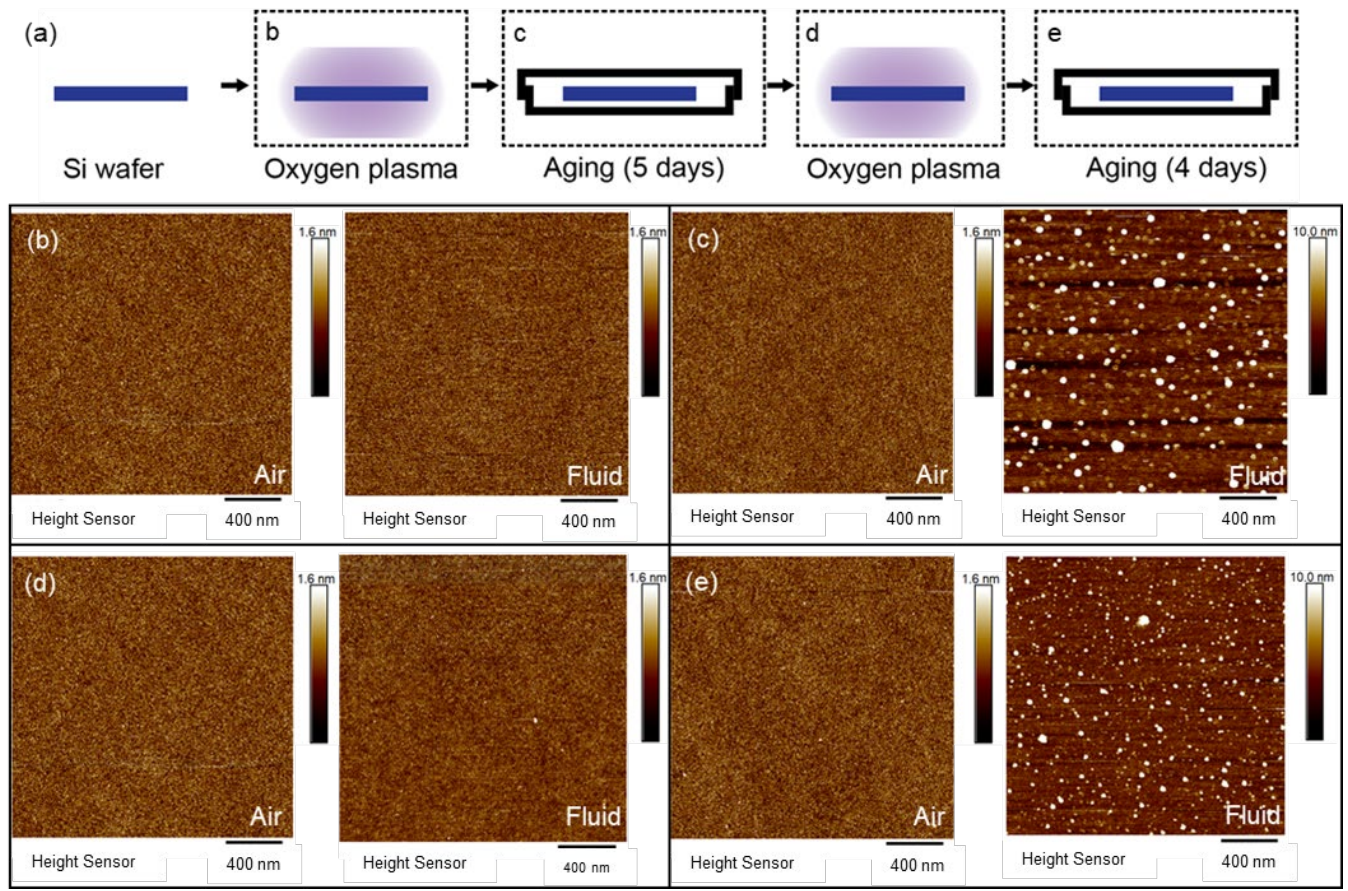

**Figure 1.** PFQNM imaging of Si surfaces in air and fluid. a) Schematic illustration of Si wafer processing steps employed to produce the samples imaged in (b)-(e). b) Plasma-treated fresh Si, with 0.1 nm RMS roughness in air and no nanobubbles present in fluid. c) Plasma-treated aged Si (5 days aging), with 0.1 nm RMS roughness in air and nanobubbles present in fluid. d) The plasma-treated aged Si surface from (c) immediately after a second plasma cleaning with 0.2 nm RMS roughness in air and no nanobubbles present in the fluid. e) Plasma-treated aged Si after the second plasma cleaning and a second aging (4 days aging), with 0.2 nm RMS roughness in air and surface nanobubbles present in fluid.

Effect of Aging Time

Figure 2 investigates the effect of aging time on the size and density of nanobubbles forming on Si. Figure 2a-d shows PFQNM images of plasma-treated aged Si surfaces exposed to DI water after 1, 5, 7, and 12 days of aging. Surface nanobubbles do not form on plasma-treated Si surfaces that are aged for 1 day (Figure 2a) but do form spontaneously and consistently on surfaces that are aged for 4 or more days (4 days = "2$^{nd}$ plasma" sample). The exact amount of time required for nanobubbles to appear on plasma-treated aged Si samples is expected to be highly sensitive to minute changes in aging conditions. For the purpose of this work, we identify >1 day as the approximate aging time required for surface nanobubbles to form on plasma-treated Si stored in a plastic wafer container in the atmospheric environment.

Figure 2e and Figure 2f provide box plots of nanobubble height ($H$) and base width ($L$), respectively, as a function of aging time. Nanobubble base width and height were measured for a minimum of 230 bubbles per sample. One-way analysis of variance was used to test for statistically significant differences in nanobubble height or base width among the four sample data sets shown in Figure 2e and Figure 2f. There are no significant differences ($p > 0.05$) in the average height or base width of nanobubbles among first plasma-treated and aged Si samples (5, 7, or 12 days) or the second plasma-treated and aged Si sample (4 days).

Figure 2g plots nanobubble density (number of bubbles per $\mu m^2$) as a function of plasma-treated Si aging time. Nanobubble densities were estimated by counting the number of bubble features identified on one 2 $\mu$m x 2 $\mu$m PFQNM image per sample. We find that prolonged exposure of the Si surface to the atmosphere does not greatly change the number density of nanobubbles on plasma-treated aged Si. While conducting PFQNM imaging, we observed that the majority of bubbles remained stable and exhibited minimal alterations even after several hours of scanning.

For all samples, nanobubbles were evenly distributed across the entire substrate. In a few instances, smaller bubbles (base width < 30 nm) vanished or formed during AFM imaging. This could be attributed to interactions with the AFM tip during scanning or the influence of neighboring bubbles.

Due to the lack of statistically significant differences in nanobubble properties for once or twice plasma-treated Si samples, and plasma-treated Si samples aged 4-12 days, the following sections do not distinguish between aging times when reporting bubble nanomechanical properties or when calculating bubble contact angle and Laplace pressure. Lumped histograms of all measured values of base width and height used in these analyses are provided in the Supporting Figure S1.

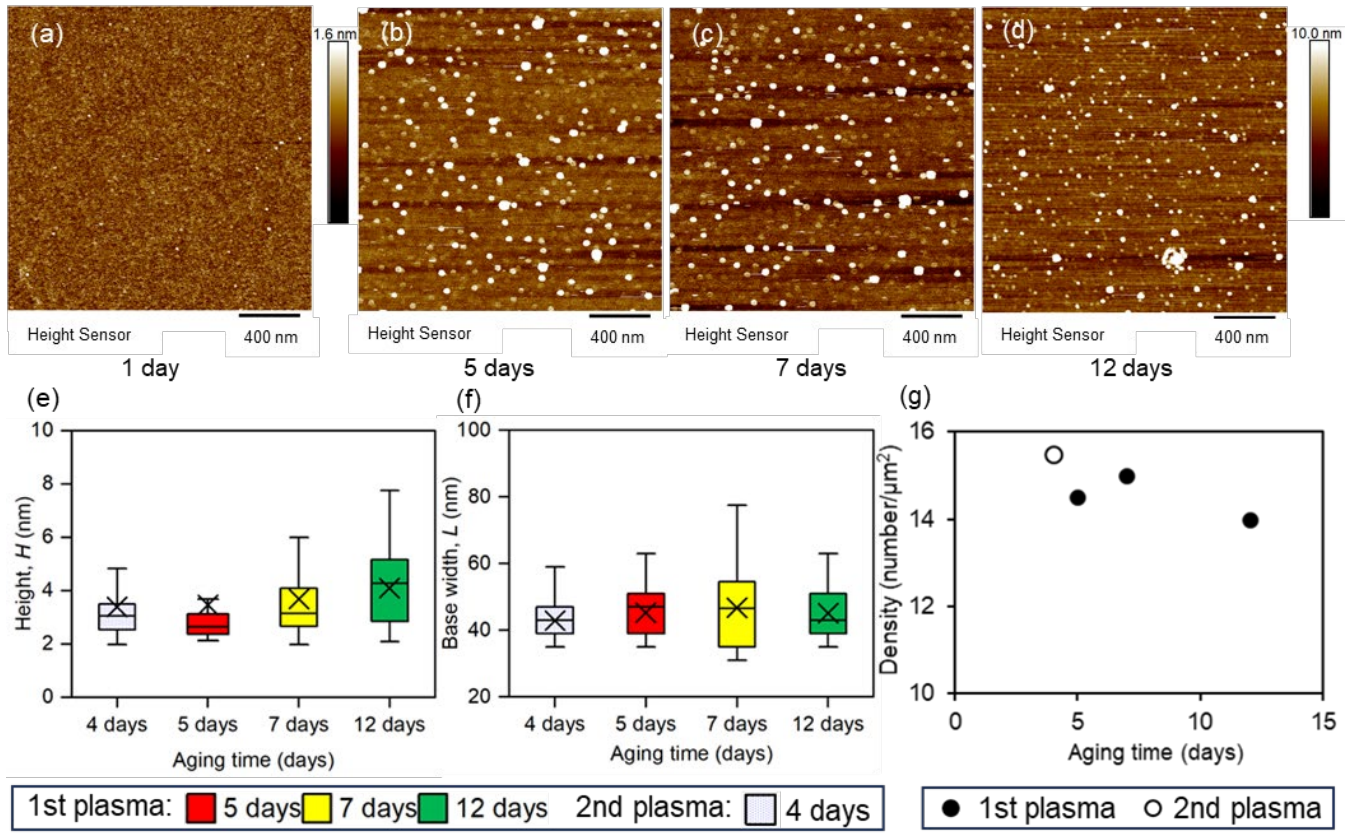

**Figure 2.** Effect of aging time on plasma-treated aged Si surface nanobubbles. a-d) PFQNM imaging (fluid) of plasma-treated Si aged for: (a) 1 day, (b) 5 days, (c) 7 days, and (d) 12 days. No bubble formation was observed on plasma-treated Si aged for 1 day. e, f) Box plots of bubble height (e) and width (f) as a function of aging time. g) Nanobubble density (number/μm$^2$) as a function of aging time, measured from 4 μm$^2$ PFQNM scans.

Plasma-Treated Aged Si Surface Nanobubble Properties

PFQNM was used to probe the mechanical properties of nanobubble features observed on plasma-treated aged Si surfaces. PFQNM imaging allows for the simultaneous capture of topographic information (Figure 3a), Derjaguin−Muller−Toporov (DMT) modulus (Figure 3b), and adhesion (Figure 3c). A comparison of plasma-treated aged Si nanobubble properties with literature values provides confidence that the observed features are surface nanobubbles comparable to those reported on other substrates.

DMT modulus mapping in Figure 3b indicates that the observed nanostructured domains exhibit considerably lower modulus than Si, supporting their identification as surface nanobubbles. Measured values of DMT modulus are sensitive to experimental factors, including tip shape, applied force, substrate-tip adhesion, point contact determination, and load-indentation response linearity. Accurate measures of DMT modulus are particularly challenging for soft materials. The results of Figure 3b are thus meant to provide a qualitative comparison of nanobubble *vs.* substrate modulus. DMT mapping in Figure 3b aligns with prior studies indicating that nanobubbles are highly deformable and very soft [13,36]. Furthermore, the modulus scale bar is similar to the scale bar representing the modulus of nanobubbles formed on aged HOPG substrate [37].

Substrate-tip adhesion mapping (Figure 3c) indicates that nanostructured domains on plasma-treated aged Si exhibit reduced adhesion compared to flat areas of the substrate. The low measured adhesion between the silicon nitride AFM tip and the nanostructured domains is consistent with previous reports of minimal adhesive forces between nanobubbles and hydrophilic AFM tips[38–40]. The presence of an attractive force between the silicon nitride tip and the more hydrophobic Si substrate in water is also theorized in literature[41] and explains the contrast in adhesive forces between nanobubbles and substrate seen in Figure 3c.

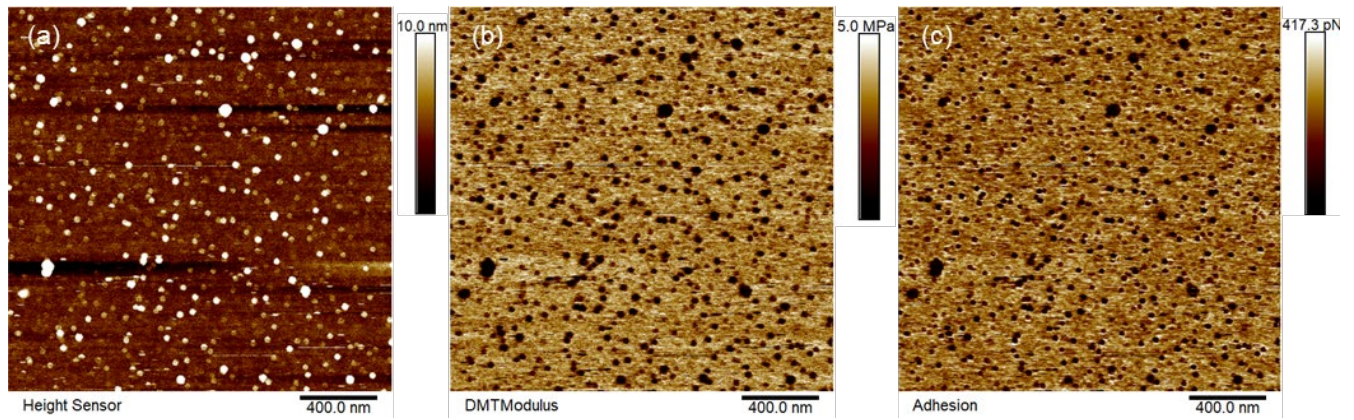

**Figure 3.** PFQNM mechanical property mapping of plasma-treated 12 days aged Si in fluid, including: (a) PFQNM imaging (topography), (b) DMT modulus, and (c) adhesion.

Further verification of plasma-treated aged Si nanostructure identity is achieved by comparing nanodomain geometric properties and Laplace pressure with literature reports for surface nanobubbles. Figure 4a defines the contact angle of a nanobubble through the gas side of the bubble and its relationship to base width ($L$), height ($H$), and radius of curvature ($R_c$). From measures of $H$ (Figure 2e) and $L$ (Figure 2f), $R_c$, is calculated using Equation 1:

$$R_c = \frac{[L^2+4H^2]}{8H} \tag{1}$$

Equation 2 determines the contact angle $\theta$ from the gas side:

$$\theta = \tan^{-1}\left[\frac{L}{2(R_c-H)}\right] \tag{2}$$

Previous reports of surface nanobubbles on hydrophobic surfaces in DI water indicate that the nanobubble contact angle is independent of base width [36]. Figure 4b plots $\theta$ vs. $L$ for 230 measured nanodomain features. The nanodomain contact angle is found to be independent of base width, as expected for surface nanobubbles. The average contact angle for all nanobubbles on plasma-treated aged Si is 17.9°.

A widely observed property of surface nanobubbles is a decrease in gas-side contact angle $\theta$ with increasing $R_c$ (equivalent to an increase in liquid-side, or "nanoscopic," contact angle with increasing $R_c$)[42]. Nanostructured domains on plasma-treated aged Si exhibit this trend, as shown in Figure 4c. The results are in good agreement with reports of van Limbeek and Seddon[42] and Song et al.[43] for surface nanobubbles nucleated on self-assembled monolayer-coated silicon substrates in contact with water.

The Laplace pressure, $\Delta P$, of a surface nanobubble is calculated using $R_c$, and the surface tension of the liquid, $\gamma$, using Equation 3:

$$\Delta P = 2\gamma/R_c \tag{3}$$

Figure 4d plots the Laplace pressures of surface nanobubbles on plasma-treated aged Si as a function of their base width ($L$). Previously, Rangharajan et al. [44] created a regime map illustrating the relationship between Laplace pressure and nanobubble base width for surfaces of varying hydrophobicity. They identified the following three regimes following an extensive literature review: 1) Regime I, $\Delta P \sim 0.01–0.1$ MPa, strongly hydrophobic substrates; 2) Regime II, $\Delta P \sim 0.1–1$ MPa, hydrophobic substrates; and 3) Regime III, $\Delta P \sim 1–10$ MPa, hydrophilic substrates. From our experimental data, the Laplace pressure of nanobubbles on plasma-treated aged Si substrates ranges from 0.9 to 3.3 MPa. This result agrees with expectations of nanobubbles on Si exhibiting properties of Regime III hydrophilic substrates. Laplace pressures of plasma-treated aged Si nanobubbles are similar to those reported for air-equilibrated water on hydrophilic self-assembled monolayers of octadecanethiol (ODT)/16-mercaptohexadecanoic acid (MHDA) on gold (Figure 4d, Song et al.)[43] and nanobubbles on hydrophilic bromo-terminated silica (Figure 4d, Rangharajan et al.) [44].

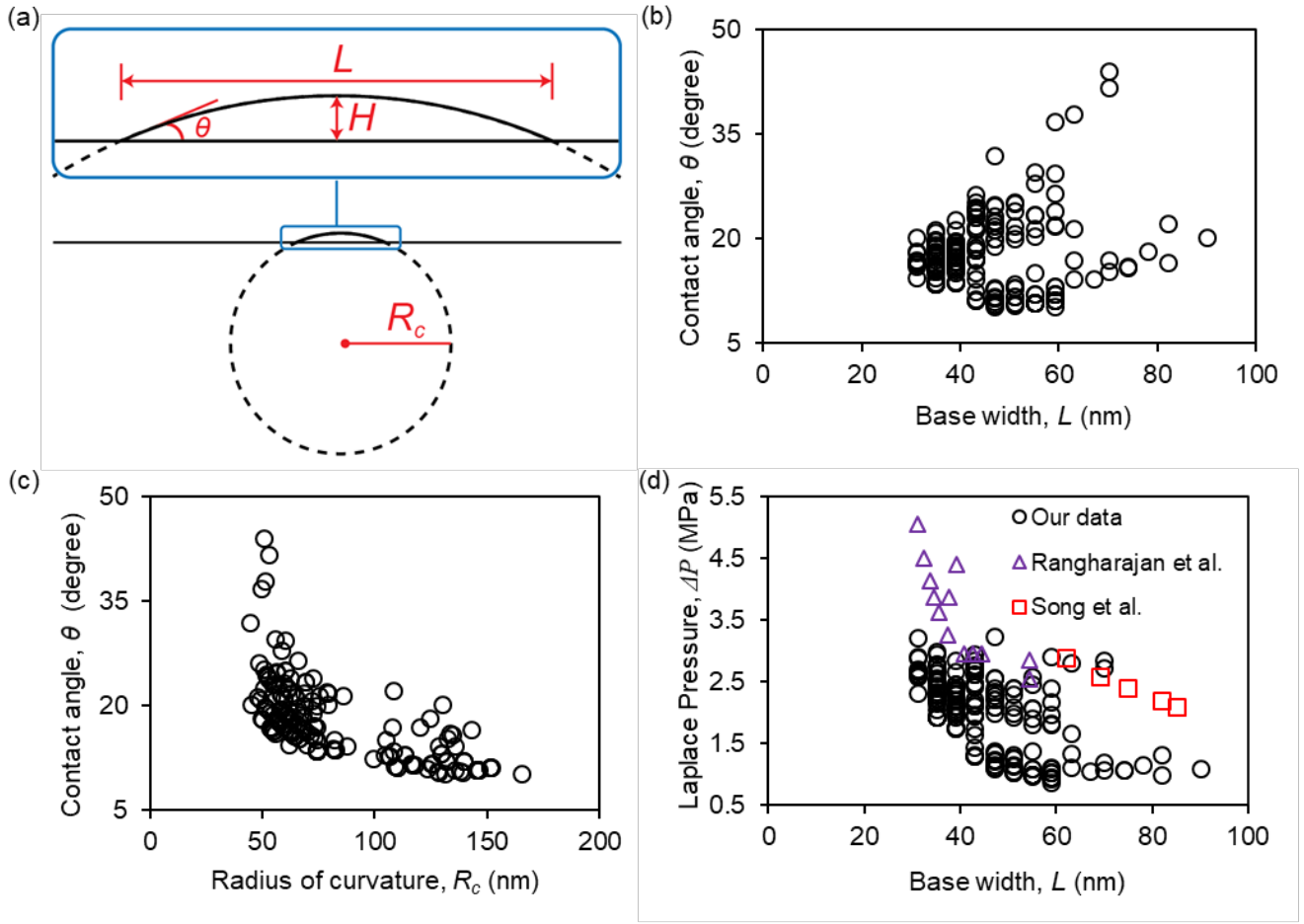

**Figure 4.** Plasma-treated aged Si surface nanobubble properties. a) Height profile of a nanobubble and schematic diagram of nanobubble geometry used for calculating relevant bubble parameters (top inset). b) Plasma-treated aged Si nanobubble contact angle, $\theta$, vs. base width, $L$. c) Plasma-treated aged Si nanobubble contact angle, $\theta$ vs. bubble radius of curvature, $R_c$. d) Laplace pressure, $\Delta P$, as a function of nanobubble base width, $L$. Laplace pressures calculated from our experimental data are in the same order of magnitude as those of Rangharajan et al.[44] and Song et al.[43].

Adventitious Carbon-Induced Bubble Formation

The presence of surface nanobubbles on plasma-treated aged Si is hypothesized to result from changes in Si surface composition and hydrophobicity with aging. More specifically, the increase in surface reactivity of Si following oxygen plasma treatment promotes adventitious carbon adsorption which increases surface hydrophobicity and enables subsequent nanobubble formation. This mechanism is supported by water contact angle and XPS measurements.

Figure 5a provides water contact angle measurements of plasma-treated Si over time. The contact angle of plasma-treated fresh Si is ~5° within one hour of plasma treatment. The contact angle increases with aging time, with the most significant increase occurring within the first 24 h after plasma treatment. Plasma-treated aged Si measures a contact angle of 27° after 1 day of aging and 38° after 7 days of aging. The water contact angle decreases to ~9° following a second oxygen plasma treatment.

Figure 5b presents XPS survey scans of plasma-treated Si after various periods of aging. During XPS measurements, samples are necessarily aged for several hours while the tool achieves the desired base pressure (typically $1 \times 10^{-10}$ Torr). Plasma-treated Si samples designated "< 1 day" aged in Figure 5 correspond to Si surfaces freshly treated by oxygen plasma and moved immediately into the XPS chamber. XPS spectra of plasma-treated fresh (<1 day) and aged Si samples show no apparent differences in C, O, Si, or F peaks present in the survey scan.

Intensities of C 1s, O 1s, Si 2s and 2p, and F 1s peaks were analyzed by XPS narrow scan and are listed in Table 1. XPS quantification accuracy is estimated at ±1%, given the very smooth surfaces and spectra collected from the wafer center position. Plasma-treated fresh Si (<1 day) exhibits a carbon content (C%) of 8.8%. Electron attenuation length calculations indicate a native oxide ($SiO_x$) thickness of approximately 1.8 nm, resulting in an oxygen content (O%) of 32.3% (Table 1). Plasma-treated aged Si surfaces have higher concentrations of C (up to 11.8%) and O (up to 38%) than plasma-treated fresh Si. The most significant changes in C and O content occur between days 1-5 of aging, consistent with water contact angle and nanobubble PFQNM measurements. Figure 5c-e provides narrow scans of C1s peaks of plasma-treated Si surfaces at high resolution. An increase in C-C and C-O peak fit intensities is observed between days 5 and 7 of aging, suggesting the possibility of longer-chain hydrocarbon adsorption with longer aging time. Additional XPS narrow scan analysis of C1s following second oxygen plasma treatment and aging (4 days) is presented in the Supporting Information (Figure S2).

The above observations are consistent with prior investigations of plasma cleaning and adventitious carbon adsorption on Si and other surfaces. Pamreddy *et al.* reported that plasma treatment increases Si hydrophilicity, reducing the nominal water contact angle of Si from 72° to <5° after 5 s of plasma cleaning[45]. Upon exposure to atmospheric conditions, Si wafers readily adsorb gaseous organic molecules from the air, even in controlled cleanroom environments. Saga *et al.* analyzed the composition of trace organic contaminants outgassing from plastic wafer boxes like the ones used in this study to store plasma-treated aged Si[46]. In addition to aromatic and aliphatic hydrocarbons, the authors detected plastic additives containing C=O and -OH groups. Such additives are especially prone to adsorb onto Si surfaces recently exposed to oxygen plasma. A change in surface wetting properties from hydrophilic to hydrophobic upon adsorption of atmospheric hydrocarbons is well documented for $Si/SiO_x$[25], Au[47], freshly cleaved HOPG [48], boron nitride nanotubes[49], and various other materials[50,51]. The ability of these hydrocarbons to accumulate over a period of days[50] and shift their composition from smaller, more volatile hydrocarbons to larger, less volatile ones during this period is also well-documented in literature[51], confirming XPS and water contact angle measurements reported in Table 1 and Figure 5.

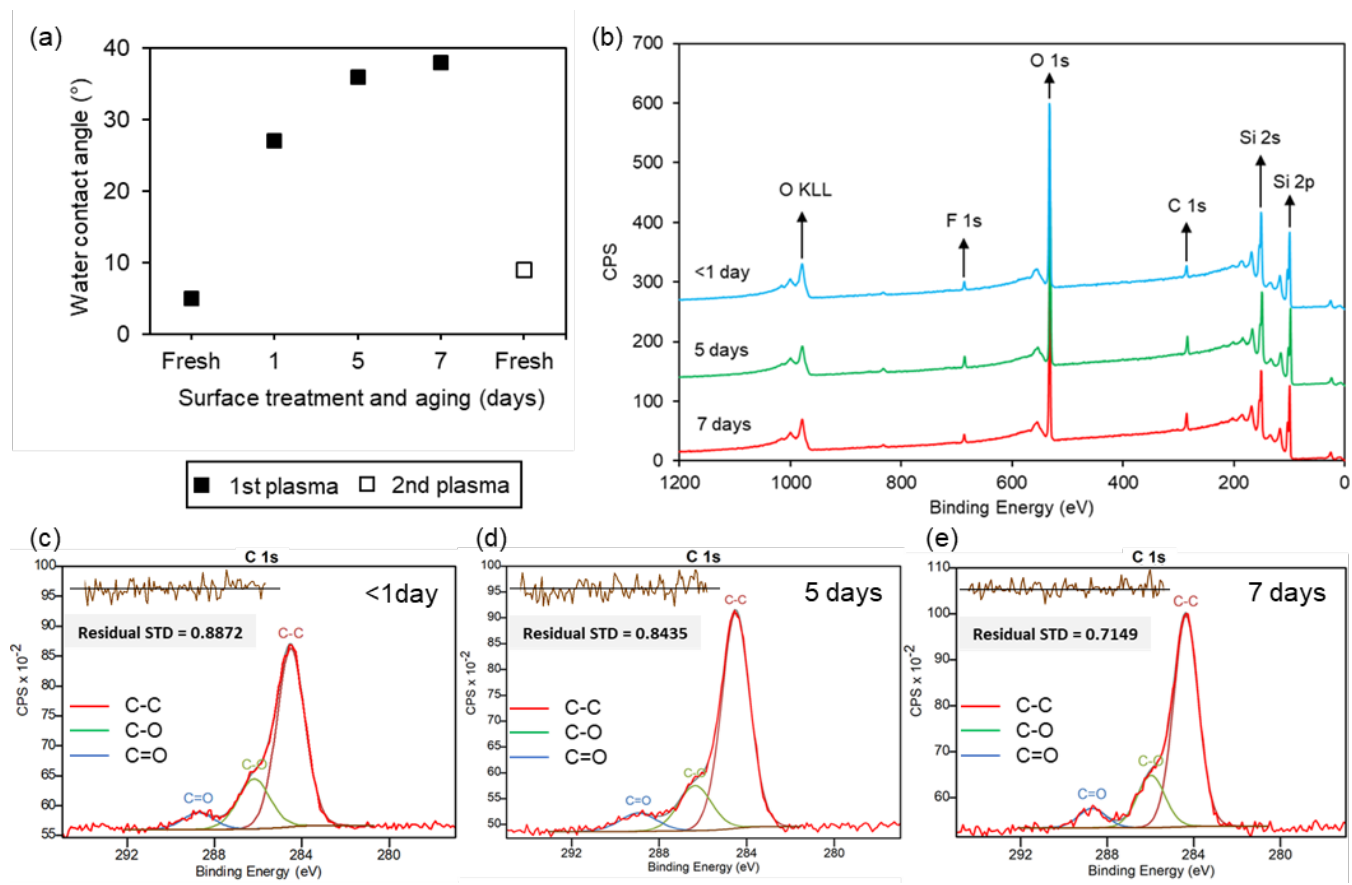

**Figure 5**. Plasma-treated aged Si surface characterization. a) Water contact angle measurements of fresh, aged (1, 5, and 7 days), and again plasma-treated Si. b) XPS survey measurements of plasma-treated fresh (< 1 day aged in the XPS tool) and plasma-treated aged (5, 7 days) Si samples. c-e) Narrow scans of C 1s peak at high resolution for fresh (< 1 day) (c), 5 day aged (d), and 7 day aged (e) samples. Peak fitting was conducted with a Shirley background and Gaussian-Lorentzian (GL(30)) synthetic line shapes. Residual STD and fit error are included in the upper left of each plot.

**Table 1.** Chemical composition of plasma-treated fresh and aged Si as determined from high-resolution XPS scans.

| Elements | Si | C | O | F | Process condition |
|---|---|---|---|---|---|
| Atomic Concentration (%) | 55.8 | 8.8 | 32.3 | 3.2 | Oxygen plasma treatment and fresh (<1 day) |
| | 50.9 | 11.2 | 35.9 | 2.1 | Oxygen plasma treatment and aged (5 days) |
| | 48.7 | 11.8 | 38.0 | 1.6 | Oxygen plasma treatment and aged (7 days) |

**Conclusion**

This study connects measurements of surface wetting change and atmospheric hydrocarbon adsorption with first-time observations of nanobubble formation on plasma-treated aged Si. Using PFQNM imaging, we observe the spontaneous formation of nanobubbles on plasma-treated aged Si, in contrast to prior work indicating that surface nanobubbles do not readily form on Si. PFQNM measurements over time reveal that > 1 day of aging in atmospheric conditions (plastic wafer container) following plasma treatment is required for surface nanobubbles to form on Si. We propose that competing adsorption takes place between smaller, volatile hydrocarbons and larger, less volatile ones, gradually changing the wetting properties of plasma-treated Si to an extent that the surface becomes highly attractive for nanobubble formation. Nanomechanical property measurements confirm the observed features as bubbles. Statistical analysis of nanobubble properties indicates that the size distribution of nanobubbles does not change significantly after plasma-treated Si surfaces age for > 1 day. The results of this work are important in all scenarios where Si surface wettability may affect adhesion and semiconductor or MEMS device fabrication.

**Supporting Information**

Histogram plots of the height and width of all nanobubbles, C 1s narrow scan of second plasma treated and aged Si.

**Author Information**

**Corresponding Author**

Roseanne Warren- *Department of Mechanical Engineering, University of Utah, 1495 E 100 S, 1550 MEK, Salt Lake City, UT 84112, USA;*

*E-mail: roseanne.warren@utah.edu

**Authors**

Anayet Ullah Siddique- *Department of Mechanical Engineering, University of Utah, 1495 E 100 S, 1550 MEK, Salt Lake City, UT 84112, USA;*

**Notes**

The authors declare no competing financial interest.

**Acknowledgments**

This work was supported in part by NSF Award #1943907 and the University of Utah Department of Mechanical Engineering. This work made use of Nanofab EMSAL shared facilities of the Micron Technology Foundation Inc. Microscopy Suite sponsored by the John and Marcia Price College of Engineering, Health Sciences Center, Office of the Vice President for Research.